\documentclass[]{spie}

\usepackage[utf8]{inputenc}
\usepackage{amsmath,amsfonts,amssymb}
\usepackage{graphicx}
\usepackage{wrapfig}
\usepackage{booktabs}
\usepackage{array}
\usepackage{tabularx}
\usepackage{enumitem}
\usepackage{placeins}
\usepackage[colorlinks=true,allcolors=blue]{hyperref}

\graphicspath{{FIGURES/}}

\newcommand{\nbar}{\overline{n}}
\newcommand{\snr}{\mathrm{SNR}}
\newcommand{\etal}{et al.}

\title{Development of quantum technologies for optical/infrared interferometry}

\author[a]{John~D.~Monnier}
\affil[a]{Department of Astronomy, University of Michigan, Ann Arbor, Michigan 48109, USA}

\authorinfo{Further author information: J.D.M.: E-mail: monnier@umich.edu}

\begin{document}
\maketitle

\begin{abstract}
Quantum technologies may revolutionize optical interferometry through new means to distribute and preserve electric-field amplitude and phase correlations between widely separated telescopes. After reviewing existing techniques, I survey proposed quantum-sensing and quantum-networking applications and highlight recent laboratory demonstrations of key building blocks. By easing field-transport demands for baselines beyond $\sim$1 km, quantum-enhanced interferometry could enable sub-milliarcsecond imaging and µas-class differential astrometry.

That said, ``going quantum'' is not a magic shortcut to sensitivity because many of the most compelling science cases remain photon- and turbulence-limited. I therefore emphasize the practical constraints -- loss, bandwidth, coherence time, synchronization, and wavelength limits of quantum interfaces -- that must also be overcome to make these capabilities useful for astronomy, whether deployed on the ground or from space.
\end{abstract}


\section{Introduction}
\label{sec:introduction}

The characteristic angular resolution of a two-aperture interferometer operating at wavelength $\lambda$ with a projected baseline $B$ is $\lambda/B$. Radio arrays routinely use continental and even Earth-sized baselines because the electric field can be amplified, down-converted, digitized, and recorded with tolerable added noise. At optical wavelengths, we normally transport the starlight itself to a common beam combiner while controlling delay, polarization, dispersion, and atmospheric phase. Existing optical/infrared arrays now extend to 330 m, providing milliarcsecond-scale resolution, and have used this approach to image stellar surfaces, circumstellar environments, stars orbiting the Milky Way's central supermassive black hole, and active galactic nuclei, as well as to make precise astrometric measurements.\cite{Monnier2003,Eisenhauer2023} Each additional reflection, meter of beam train, and increment of delay-line stroke introduces loss or instability or expense, however, and a lost astronomical photon cannot be replaced. These constraints become increasingly important for kilometer-scale and longer baselines.

Quantum information science suggests another architecture. The astronomical field remains local at each telescope while an engineered quantum resource, such as a reference photon, an entangled pair shared between stations, or a state held in memory, is distributed or prepared through a network. Measurements made locally at the telescopes are then compared through an ordinary classical communications channel. In principle, this can recover the first-order coherence measured by a conventional beam combiner without transporting starlight over the full baseline. Unlike the unknown astronomical field, the engineered resource can be generated repeatedly, its successful preparation can be confirmed (``heralded''), and it can sometimes be stored until needed. Gottesman, Jennewein, and Croke introduced this approach for telescope arrays using quantum repeaters, which are intermediate nodes that establish entanglement over a long link by connecting shorter links.\cite{Gottesman2012} Later work has explored entanglement consumption, time-bin encoding, quantum memories, improved circuits, larger arrays, and laboratory tests.\cite{Khabiboulline2019PRL,Khabiboulline2019PRA,Czupryniak2023,Brown2023,Crawford2023,Stas2026} While these parallel approaches may favor different bandwidths, storage times, link lengths, and source brightnesses, the ideal lossless direct detection remains the benchmark.

This paper follows my 2026 SPIE review talk and is not an encyclopedic survey. I write from the perspective of an astronomer and instrument builder. I first discuss photons per mode, then compare classical and quantum-assisted measurements. I next review representative protocols and experiments, and close with observatory requirements and possible next tests. Complementary recent reviews and roadmaps are provided by Huang \etal\ and Rajagopal \etal\cite{Huang2026,Rajagopal2024}.

\section{Why optical interferometry is photon-starved}
\label{sec:photonstarved}

\subsection{Photons per mode, not photons per second}
The most critical quantity for weak-light interferometry is the mean photon occupation per mode, $\nbar$, not the total photon rate. Here a mode means one independently measurable optical-field degree of freedom, with a specified spatial pattern, polarization, and temporal or spectral waveform. A bright star can supply many photons per second to a telescope, but if those photons are spread over many independent modes, each mode may be nearly empty.

For one polarization mode of thermal radiation at frequency $\nu$ and temperature $T$,
\begin{equation}
    \nbar(\nu,T)=\frac{1}{\exp(h\nu/k_{\rm B}T)-1}.
    \label{eq:occupation}
\end{equation}
A single frequency-temporal mode obeys the condition that the product of its bandwidth $\Delta\nu$ and its coherence time $\tau_c$ is of order unity, $\Delta\nu\,\tau_c\sim1$. A single spatial mode is diffraction limited, with angular area $\Omega_{\rm mode}\sim(\lambda/D)^2$ for a telescope of diameter $D$.
Increasing bandwidth collects more photons by adding independent spectral modes, not by making each mode brighter. Exploiting a broad band therefore requires the instrument to process many modes in parallel. During an integration $\tau$, the photons are spread over approximately
\begin{equation}
    M \simeq N_{\rm pol}\,\Delta\nu\,\tau
    \label{eq:modes}
\end{equation}
independent modes, up to order-unity conventions for the temporal coherence interval. Here $N_{\rm pol}$ is the number of polarization states included. There is also a spatial-dilution penalty for an unresolved star: it fills only a small fraction of the telescope's diffraction-limited spatial mode. Schematically,
\begin{equation}
    \nbar_{\rm eff}\simeq \nbar\,
    \min\!\left[1,\frac{\Omega_{\rm source}}{\Omega_{\rm mode}}\right],
    \qquad \Omega_{\rm mode}\sim\left(\frac{\lambda}{D}\right)^2.
    \label{eq:effectiveoccupation}
\end{equation}
This expression includes source spatial dilution but not instrument coupling or downstream losses, which are included later in $\eta_q$.

Table~\ref{tab:occupation} gives examples from the review talk. At $1~\mu$m, the illustrative 10th-magnitude star supplies only $\nbar_{\rm eff}\simeq7.5\times10^{-9}$ to a diffraction-limited 1-m aperture and $7.5\times10^{-7}$ to a diffraction-limited 10-m aperture. Its angular area fills only a small fraction of the telescope's spatial mode. A larger aperture improves this filling factor but cannot increase the surface brightness above the Planck occupation.

\begin{table}[ht]
\caption{Illustrative mean photon occupations per thermal mode. The effective values include spatial dilution for the stated telescope diameter (assuming diffraction-limited performance) but not a detailed instrument throughput.}
\label{tab:occupation}
\centering
\small
\begin{tabular}{lccc}
\toprule
Source and wavelength & $\nbar$ & $\nbar_{\rm eff}$ (D$=$1\,m) & $\nbar_{\rm eff}$ (D$=$10\,m) \\
\midrule
Sun (6000 K) at $1~\mu$m            & $1.0\times10^{-1}$ & $1.0\times10^{-1}$ & $1.0\times10^{-1}$ \\
\addlinespace[2pt]
Betelgeuse (3000 K) at $1~\mu$m     & $8.3\times10^{-3}$ & $3.1\times10^{-4}$ & $8.3\times10^{-3}$ \\
\addlinespace[2pt]
10th-mag star (10000 K) at $1~\mu$m & $3.1\times10^{-1}$ & $7.5\times10^{-9}$ & $7.5\times10^{-7}$ \\
\addlinespace[2pt]
Thermal source (300 K) at 1 cm      & $2.1\times10^{2}$  & $2.1\times10^{2}$  & $2.1\times10^{2}$ \\
\addlinespace[2pt]
Thermal source (5000 K) at 1 cm     & $3.5\times10^{3}$  & $3.5\times10^{3}$  & $3.5\times10^{3}$ \\
\bottomrule
\end{tabular}
\end{table}

\subsection{What the classical architectures measure}

Before considering new quantum approaches, I briefly review three classical architectures for stellar interferometry: direct detection, Hanbury Brown--Twiss (HBT) intensity interferometry, and heterodyne detection. The direct-versus-heterodyne trade has also been examined explicitly for a large, many-baseline astronomy facility in the Planet Formation Imager concept.\cite{irelandmonnier2014}

Direct detection brings the optical fields collected by the telescopes together in a common beam combiner before photodetection. Let $E_A$ and $E_B$ be the complex fields coupled into one matched mode at telescopes A and B. Their complex degree of first-order coherence is the visibility $V_{AB}$:
\begin{equation}
    V_{AB}=\frac{\langle E_A^*E_B\rangle}
    {\sqrt{\langle |E_A|^2\rangle\langle |E_B|^2\rangle}}.
    \label{eq:visibility}
\end{equation}
The van Cittert--Zernike theorem identifies $V_{AB}$ with a Fourier component of the sky brightness. A conventional beam combiner superposes the two optical paths and measures this first-order coherence. With lossless transport and ideal detectors, direct detection is broadband and has the best weak-light scaling.

Hanbury Brown--Twiss (HBT) intensity interferometry instead detects photons independently at each telescope and correlates their intensity fluctuations. For thermal light, the normalized zero-delay intensity correlation is
\begin{equation}
    g^{(2)}_{AB}(0)=1+|V_{AB}|^2.
    \label{eq:hbt}
\end{equation}
HBT does not require a phase-stable optical link. Its path-length tolerance is set mainly by detector timing rather than by optical-phase matching, although coarse delay must still keep the signals within the correlation window. Because HBT does not require diffraction-limited single-mode coupling, it can use large, relatively inexpensive ``light bucket'' telescopes and relax adaptive-optics requirements. The price is a weak second-order signal when $\nbar\ll1$, measured against many accidental coincidences.\cite{HanburyBrown1956,Tsang2011} See Table~\ref{tab:scaling} for the weak-light scaling of HBT.

In heterodyne detection, each telescope mixes the source with a phase-referenced local oscillator (LO) and records the beat signal locally. A major advantage is that the LO can remain coherent over long times, transferring the optical phase into a lower-frequency electronic record; the LOs at different stations must themselves remain mutually phase referenced. Those electronic records can be copied with little additional noise, then delayed and correlated in software. This is especially useful for arrays with many telescopes because the same recorded signal can feed many baseline combinations.\cite{irelandmonnier2014} The price is that even an ideal heterodyne receiver has a quantum noise floor equivalent to roughly one photon per measured mode. Because optical starlight typically has \(\bar n\ll1\) photon per mode, this noise overwhelms the signal.

Table~\ref{tab:scaling} gives the leading weak-light scalings, omitting order-unity constants and estimator details. Here $M$ is the number of incident matched modes used in the comparison, $|V|$ is the source visibility amplitude, and $\eta_q$ is an architecture-specific effective efficiency that includes the relevant throughput and contrast losses. The same symbol is used in every row to keep this penalty explicit, but it should not be interpreted as the same component-level efficiency for every architecture. For the quantum-assisted case, $\eta_q$ is the usable distributed-resource factor expanded in Eq.~\ref{eq:etaq}. Direct detection and entanglement-assisted interferometry therefore have square-root throughput dependence, while heterodyne and HBT have linear throughput dependence in the weak-light limit. Ideal first-order detection scales as $\sqrt{\nbar}$; HBT and heterodyne scale as $\nbar$ when $\nbar\ll1$.

\begin{table}[ht]
\caption{Schematic weak-thermal-light sensitivity scalings. These expressions compare idealized architectures and omit order-unity factors, backgrounds, detector dark counts, and calibration systematics.}

\label{tab:scaling}
\centering
\small
\begin{tabularx}{\linewidth}{>{\raggedright\arraybackslash}p{0.19\linewidth} >{\centering\arraybackslash}p{0.23\linewidth} >{\raggedright\arraybackslash}X}
\toprule
Architecture & Representative scaling & Dominant practical trade \\
\midrule
Direct detection & $\snr\propto |V|\sqrt{\eta_q M\nbar}$ & Optimal use of weak light, but the astronomical field must reach a common combiner. \\
\addlinespace[2pt]
Heterodyne & $\snr\propto |V|\eta_q\nbar\sqrt{M}$ & Local measurement and electronic correlation, but vacuum/local-oscillator noise dominates when $\nbar\ll1$. \\
\addlinespace[2pt]
HBT intensity & $\snr\propto |V|^2\eta_q\nbar\sqrt{M}$ & Phase-insensitive long-baseline operation, but measures a weak second-order correlation. \\
\addlinespace[2pt]
Entanglement assisted & $\snr\propto |V|\sqrt{\eta_q M\nbar}$ & Can recover direct-like scaling only if entanglement, local interactions, and detection are sufficiently efficient. \\
\bottomrule
\end{tabularx}
\end{table}

\begin{figure}[ht]
\centering
\includegraphics[width=0.98\linewidth]{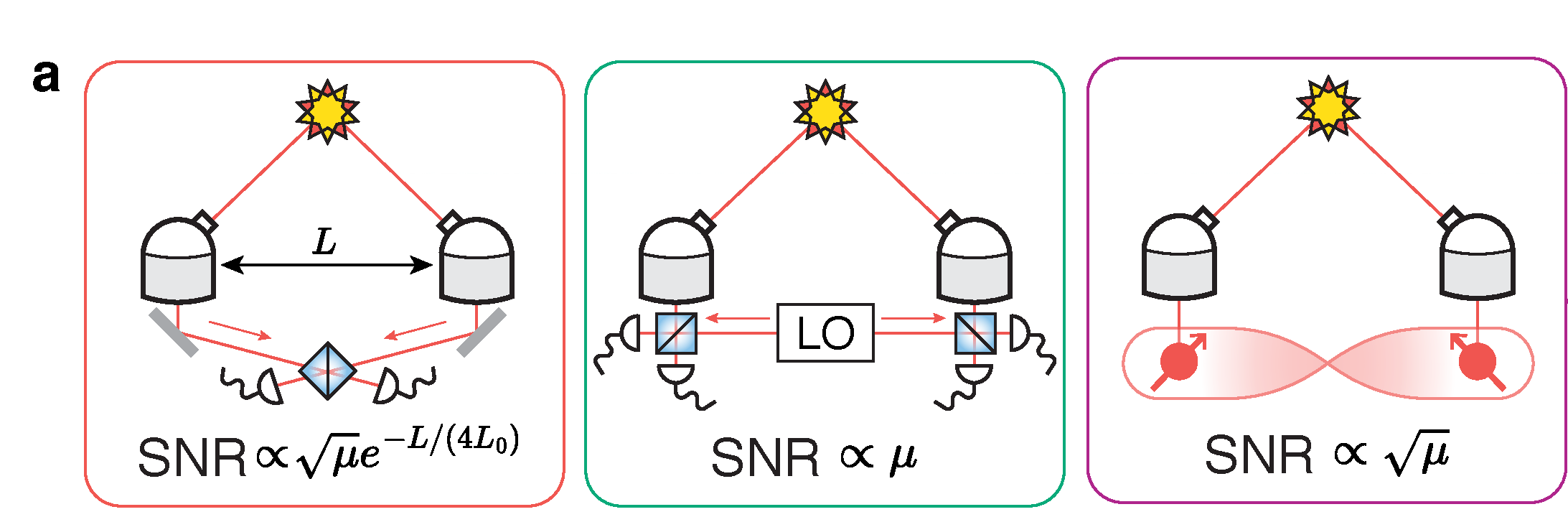}
\caption{Three phase-sensitive architectures and representative weak-source SNR scalings; the source's $\mu$ corresponds to $\nbar$ in our scalings. Cropped from Fig.~1a of P.-J. Stas \etal, arXiv:2509.09464v2 (2025), under the \href{https://creativecommons.org/licenses/by/4.0/}{CC BY 4.0} license; see also the published version.\cite{Stas2026}}
\label{fig:architectures}
\end{figure}

Figure~\ref{fig:architectures} compares three phase-sensitive architectures that access complex first-order coherence: direct transport, local-oscillator detection, and quantum-assisted nonlocal detection. HBT is deliberately absent because it instead correlates locally measured intensities to obtain the second-order quantity $|V|^2$. Quantum-assisted performance depends on the efficiency, fidelity, and rate of the distributed resource.

\FloatBarrier
\section{What quantum interferometry can and cannot do}
\label{sec:guardrails}

For a quantum-assisted interferometer to offer advantages of direct-detection methods, it would have to:
\begin{enumerate}[leftmargin=*,itemsep=2pt]
    \item keep the astronomical photons local to each telescope;
    \item distribute entanglement, reference photons, or stored quantum states through a network designed for that purpose;
    \item open baselines far beyond the kilometer scale without building a kilometer-scale optical delay line; and
    \item retain the first-order-coherence sensitivity that makes direct detection attractive.
\end{enumerate}
Detectors, memories, frequency conversion, stabilized links, and clocks have advanced enough to test several required operations together. The main question is now whether a complete system can process enough astronomical modes at a useful rate.

The quantum toolkit does not remove several familiar physical limits. The main guardrails are:

\begin{itemize}[leftmargin=*,itemsep=2pt]
    \item \textbf{No free copies and no noiseless amplifier.} A quantum memory can store a captured state and return it later, but it cannot make a perfect spare copy. Nor can we noiselessly amplify an arbitrary faint field. If the photon is lost before local capture, quantum processing cannot bring it back.

    \item \textbf{Teleportation is real, but constrained.} It can transfer an unknown quantum state, but only by consuming entanglement already shared between the stations and sending the result of a local measurement through a classical channel. Nothing travels faster than light, and causality is not violated. For an astronomical interferometer, this means the entangled resource must be distributed and ready when the signal mode is processed.

    \item \textbf{Which-path information kills the phase.} If the measurement reveals which telescope received the photon, the interference is lost. The protocol may herald that a photon arrived, but it must hide which path the photon took (``path erasure''). The same requirement appears in the earlier quantum-assisted telescope-array protocol, and the recent experimental literature calls the corresponding operation ``photon mode erasure.''\cite{Khabiboulline2019PRA,Stas2026}

    \item \textbf{There is no software delay line after detection.} Heterodyne uses a long-coherence LO to transfer the optical phase into an electronic record that can be aligned later in software. Flying-reference protocols such as GJC have no such electronic record: the astronomical and reference photons must overlap when they are combined locally. Memory-assisted protocols can instead retain the relevant phase in a stationary quantum state, but the stellar mode must be coupled into the memory at the correct time and the write, storage, and readout phases must be calibrated. In either case, post-processing cannot restore coherence that was lost before the measurement.

    \item \textbf{The observatory problems remain.} Quantum hardware does not remove atmospheric piston, chromatic dispersion, variable coupling, background, or source variability. We still need collecting area, fringe tracking to stabilize atmospheric piston during an exposure, delay tracking to keep the optical paths within the coherence envelope, and calibration.
\end{itemize}

I first describe the principal protocols and their theoretical extensions, and then turn to what the laboratory experiments actually demonstrated.

\section{Protocols and theory}
\label{sec:protocols}

\subsection{GJC: flying path-entangled reference photons}

The Gottesman--Jennewein--Croke (GJC) protocol is designed to let two telescopes measure the same complex visibility as a direct beam combiner without transporting the starlight to a common location. Here, the starlight remains at each telescope and interacts with a single distributed reference photon placed in a coherent superposition of traveling to telescope A or telescope B. The detector outcomes recorded at the two stations are combined through a classical channel. A controllable phase in the reference allows these correlations to recover both the visibility amplitude and phase.\cite{Gottesman2012}

In its simplest form, the path-entangled reference photon is
\begin{equation}
    |\psi_{\rm ref}\rangle = \frac{1}{\sqrt{2}}
    \left(|1\rangle_A|0\rangle_B + e^{i\delta}|0\rangle_A|1\rangle_B\right).
    \label{eq:reference}
\end{equation}
Here $|1\rangle_A|0\rangle_B$ means one reference photon at telescope A and none at telescope B, while $|0\rangle_A|1\rangle_B$ means the reverse; $\delta$ is their tunable relative phase. Each telescope combines its astronomical mode and its local share of the reference on a beam splitter. When one photon is detected at each station, the two possibilities cannot be distinguished: a stellar photon at $A$ and a reference photon at $B$, or vice versa. The balance between correlated and anticorrelated detector outcomes then measures one phase projection of the visibility, proportional to $\mathrm{Re}(V_{AB}e^{-i\delta})$. Varying $\delta$ samples different projections.

Spontaneous parametric down-conversion (SPDC) is a common laboratory source of correlated photon pairs. A pump laser illuminates a nonlinear crystal, where occasionally one pump photon is converted into two lower-energy photons. One photon of the pair is detected locally as a herald. That click tells us that its partner was probably produced and identifies the time bin to keep. The partner, which is the reference photon used by the interferometer, is sent through a beam splitter. It is not directed to one telescope or the other. Instead, the beam splitter puts it into a coherent superposition of the two paths, and the two outputs are sent to the two telescope stations. This is the path-entangled reference in Eq.~\ref{eq:reference}.

Creation of the reference photon is not the only challenge, as the timing requirements for distribution are unforgiving. After the reference photon is divided at the beam splitter, the differential propagation time to the two stations must track the differential arrival time of the stellar wavefront at the telescopes. The astronomical and reference photons must overlap within the same time bin (of order 1 nanosecond) when they are combined locally, as well as being matched in spectrum, polarization, and spatial mode. While this is fundamentally the same challenge direct-detection interferometrists must solve when bringing starlight together for interference, there are some advantages to distributing the reference photon rather than the stellar photon. The reference is an engineered resource: its emission time, spectrum, polarization, spatial mode, and phase can be controlled, and a future network could use heralding, wavelength conversion, or quantum repeaters to establish it over long distances. A lost reference can be replaced for a later trial, whereas starlight lost in a long beam train is irretrievable. The astronomical photon also remains in a short local path from the telescope to the detector, potentially maintaining high quantum efficiency for this most precious resource. These advantages do not relax the final requirement: a successfully delivered reference photon must still be present and mode matched when the stellar photon arrives.

Note that heterodyne methods avoid this pre-detection optical overlap by using a long-coherence LO to preserve the phase in a local electronic record; the flying-reference measurement does not leave such a record. There is therefore no possibility of a software delay line after photodetection, as there is for HBT or heterodyne methods. We discuss later how a quantum memory could instead provide a physical, programmable delay, but the storage interval, readout timing, and phase must still be set and calibrated before the final measurement; the memory moves the delay-matching step rather than eliminating it. One hopes the promised quantum network can distribute resources with the required timing. Otherwise, the future quantum interferometrist will still need to build the correct delay and phase control into a bespoke network, reducing some of the practical advantage of the quantum approach.

The GJC protocol has two important disadvantages relative to direct detection and the memory-assisted Khabiboulline protocol we discuss next. First, in the ideal balanced two-station scheme, the astronomical and reference photons arrive at the same station in half of the trials, and these events must be discarded. This halves the usable astronomical-photon throughput, contributing an intrinsic factor of $1/2$ to the end-to-end efficiency $\eta_q$ before other losses. Second, because stellar-photon arrival times are random, a mode-matched flying reference photon must be available in every temporal mode. For 1-ns time bins, this implies a reference-photon rate of order 1~GHz per spectral, spatial, and polarization channel. Producing such photons at high efficiency and with low multiphoton contamination, distributing them through a quantum network, and maintaining adequate mode matching remain challenging. If a stellar photon arrives in a mode without a matching reference photon, that event is lost.\cite{Gottesman2012}

\subsection{Khabiboulline: pre-shared Bell pairs and quantum memories}

The Khabiboulline protocol is designed to let two telescopes measure the same complex visibility as a direct beam combiner without transporting the starlight or supplying a flying reference photon in every possible time bin. As for GJC, the starlight again remains at each telescope. Unlike GJC, each telescope couples its incoming light to local quantum memories, while Bell pairs are distributed between the stations in advance. A Bell pair consists of one two-state memory element, or qubit, at each station, prepared in a shared quantum state whose measurement outcomes remain correlated. Local operations on the memories and Bell pairs preserve the possibility that the photon arrived at either telescope (``path erasure'' concept), and the results from the two stations are combined through a classical channel to recover both the visibility amplitude and phase.\cite{Khabiboulline2019PRL,Khabiboulline2019PRA}

The key practical attraction is that these Bell pairs can be prepared and stored before the unpredictable stellar photon arrives. Useful stellar photons in a given measured mode arrive slowly, whereas the network can make many heralded attempts to establish Bell pairs between typical photon arrivals. An individual attempt need not succeed: Bell-pair generation can be retried, provided the complete entangled resource is ready before the next stellar photon arrives and remains coherent until it is used. This retry advantage does not extend to the signal: every missed stellar photon is an irrecoverable loss that reduces the accumulated SNR.

In more detail, every incoming time bin interacts with a local memory. If a stellar photon is present, its optical state and arrival-time address are written into memory. The remaining optical mode is then measured in a way that erases which-telescope information, an operation called ``photon-mode erasure'' in recent experimental work. The shared Bell pairs are used in coordinated same-or-different tests, called parity checks, to identify the occupied time bin without determining which telescope received the photon. The protocol therefore learns \emph{when} the photon arrived while deliberately leaving \emph{where} it arrived unresolved, preserving the coherence between ``photon at telescope A'' and ``photon at telescope B.'' The memories are finally measured using tunable local phases, and the joint statistics recover the real and imaginary parts of the visibility. Those local phases must still be tied to a stable correlated reference, such as phase-linked control fields; the Bell pairs alone do not supply it.\cite{Khabiboulline2019PRL,ZhangJennewein2025,Stas2026}

In the ideal protocol, all astronomical-photon events can contribute rather than half being discarded as in GJC, and no mode-matched reference photon is required in every possible time bin. Binary encoding keeps the prepared resource modest: a block of $M_{\rm bin}$ possible arrival-time bins requires only $\lceil\log_2(M_{\rm bin}+1)\rceil$ stored Bell pairs rather than one pair per bin. The demanding part shifts to the local light--memory interface, which must still process every incoming time bin and capture the stellar state when it appears, as well as to memory coherence and low-error quantum gates. Later circuit designs reduced the supporting qubits and quantum gates further.\cite{Khabiboulline2019PRL,Khabiboulline2019PRA,Czupryniak2023} Unlike GJC, this protocol does not require an astronomical photon to overlap locally with a flying reference photon. It still requires precise timing: the geometric delay and station clocks must be known well enough that corresponding stellar temporal modes are assigned to the same memory time bin at the two telescopes, within the coherence time and with their relative phase preserved.

\subsection{More than two telescopes}

A single baseline is not an imaging array. Two telescopes provide one complex visibility, whereas useful imaging normally requires many baselines and robust quantities such as closure phase.\cite{Jennison1958,MonnierPhases} For three stations,
\begin{equation}
    \Phi_{123}=\arg\left(V_{12}V_{23}V_{31}\right),
    \label{eq:closure}
\end{equation}
which cancels telescope-based phase errors in an ideal classical array. Extending a quantum-assisted protocol from one baseline to an imaging array is not automatic. GJC suggested sharing one reference photon coherently among all telescopes so that reference errors could cancel in closure combinations, while Khabiboulline \etal\ extended their memory architecture to $N$ stations using multi-station entanglement.\cite{Gottesman2012,Khabiboulline2019PRA} These proposals showed how to connect more than two telescopes; the explicit construction of a robust quantum closure-phase measurement has been addressed only more recently.

Multi-telescope quantum interferometry is an especially active topic, with several complementary theoretical results appearing in 2026. For conventional imaging, pairwise complex visibilities contain the needed image information. The open question is whether measuring each pair separately is quantum-optimal, or whether quantum states and joint measurements shared across $N>2$ telescopes can extract that information more efficiently. An SPIE contribution by Gorshkov \etal, \textit{Quantum-optimized phase closure for robust imaging}, treats the closure observable directly.\cite{Gorshkov2026PhaseClosure} Padilla, Sajjad, Saif, and Guha showed how local spatial-mode sorting and entanglement-assisted operations could synthesize an arbitrary linear-optical measurement across any number of telescopes; their detailed example uses two telescopes, so the larger-array result is a general receiver construction rather than a derived closure observable.\cite{Padilla2026} A 2026 preprint by Zhang and collaborators gives a weak-light imaging example in which genuinely multi-telescope measurements can outperform independent pairwise receivers for some image parameters.\cite{ZhangMultiTelescope2026} Together, these works address robust phase closure, general array-receiver design, and the possible need for genuinely multi-telescope measurements. The Kwiat group's conference sequence, discussed below, began with a three-arm layout and one-baseline fringe data, then reported visibility-amplitude recovery under imposed turbulence and a preliminary zero-closure result for a point source.\cite{Diaz2021,Diaz2022,Diaz2024Poster}

\section{Selected experiments and proposals}
\label{sec:experiments}

The previous section described the protocols and theoretical extensions; I now ask what the laboratory experiments actually measured. The protocol correspondence is important: Brown \etal\ implemented the GJC entangled-reference idea with real optical hardware; Stas \etal\ demonstrated Khabiboulline-style entanglement-assisted nonlocal phase sensing between two quantum-memory nodes, although not its binary time-bin encoding; and Wang \etal\ implemented a memory-assisted Brown/GJC receiver with programmable delay compensation.

\subsection{Brown--Smith entangled-reference experiment}

\begin{figure}[ht]
\centering
\includegraphics[width=0.98\linewidth]{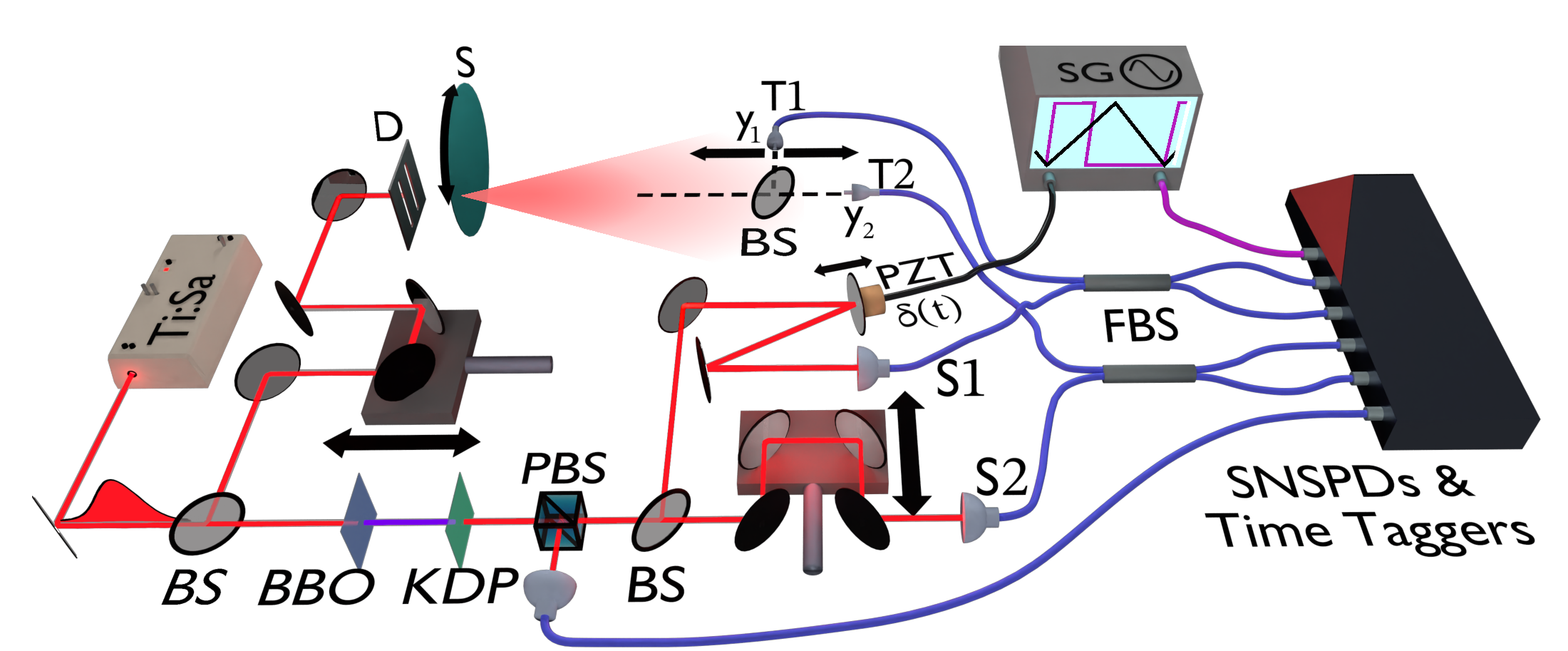}
\caption{Laboratory implementation of the entangled-reference receiver of Brown \etal. A rotating diffuser produced the weak pseudo-thermal source, while SPDC supplied the heralded path-entangled reference. Cropped from Fig.~2 of M. R. Brown \etal, ``Interferometric Imaging Using Shared Quantum Entanglement,'' \textit{Phys. Rev. Lett.} \textbf{131}, 210801 (2023), \href{https://doi.org/10.1103/PhysRevLett.131.210801}{doi:10.1103/PhysRevLett.131.210801}, under the \href{https://creativecommons.org/licenses/by/4.0/}{CC BY 4.0} license; cropping is the only modification.}
\label{fig:gjc}
\end{figure}

Brown, Smith, and collaborators at the University of Oregon implemented the GJC entangled-reference idea with real optical hardware in a tabletop experiment.\cite{Brown2023} A laser scattered from a rotating diffuser supplied pseudo-thermal ``starlight,'' meaning laboratory light with thermal-like intensity fluctuations, while an SPDC source produced the photon pairs. Detection of one photon heralded the event; its partner passed through a beam splitter to form the path-entangled reference sent toward the two receiver arms. At each arm, the reference was combined with the local pseudo-starlight, and superconducting nanowire single-photon detectors recorded the coincidences. Figure~\ref{fig:gjc} shows the apparatus.

The experiment used two double-slit configurations of the same pseudo-thermal source as binary-star analogues. At each baseline setting, it recovered a complex visibility phasor, including amplitude and phase. Because the phase reference was not stable between different baseline settings, those phasors could not be combined directly into a phase-referenced image. The analysis therefore used $|V|^2$ across baselines and recovered the source-intensity autocorrelation. The herald detector was a threshold device: it distinguished a pair-production event from an empty time bin, but could not certify that exactly one pair had been produced, so occasional multipair events could pass the herald (something that could be avoided with a photon-number-resolving detector). The apparatus included a heralded-photon source, detectors, and real mode-matching errors, but not separated telescopes, atmospheric propagation, or an astronomical source. Spatially filtered sunlight is one possible next test.

\subsection{Event-ready diamond-memory nodes}

Stas \etal\ demonstrated entanglement-assisted nonlocal optical phase sensing using two separated quantum-memory nodes, implementing the central two-station measurement of the Khabiboulline approach rather than the GJC flying-reference receiver.\cite{Stas2026} They first generated and heralded a Bell pair between nuclear-spin memories at the two stations. Here ``event-ready'' means that this shared entangled resource was confirmed before the weak signal was measured. A weak phase-encoded optical signal then interacted locally with each diamond node. Additional local optical fields erased which-station information, so a detector click did not reveal where the signal photon had arrived. The electron spin at each node was measured locally, and the two outcomes were compared to flag ``signal present somewhere'' without revealing which node received the signal. The nuclear-spin memories were then measured locally in phase-sensitive bases, and correlations between those outcomes recovered the differential phase. Thus the experiment combined remote Bell-pair distribution, local light--memory interaction, photon mode erasure, nonlocal photon heralding, and phase-sensitive interferometric readout. It did not implement Khabiboulline's binary encoding of many possible arrival-time bins.

Each node used a silicon-vacancy center, an atom-like defect in diamond whose electron spin provided the fast interface to light. A nearby nuclear spin provided the longer-lived memory; both spins were manipulated with microwave pulses that drove the quantum gates and rotations. The signal-present test accepted only a fraction of real photon events, and imperfections in the shared Bell pair also produced false triggers, equivalent to reporting a photon when none was present. The input was a phase-scrambled attenuated laser rather than starlight.

\begin{figure}[ht]
\centering
\includegraphics[width=0.98\linewidth]{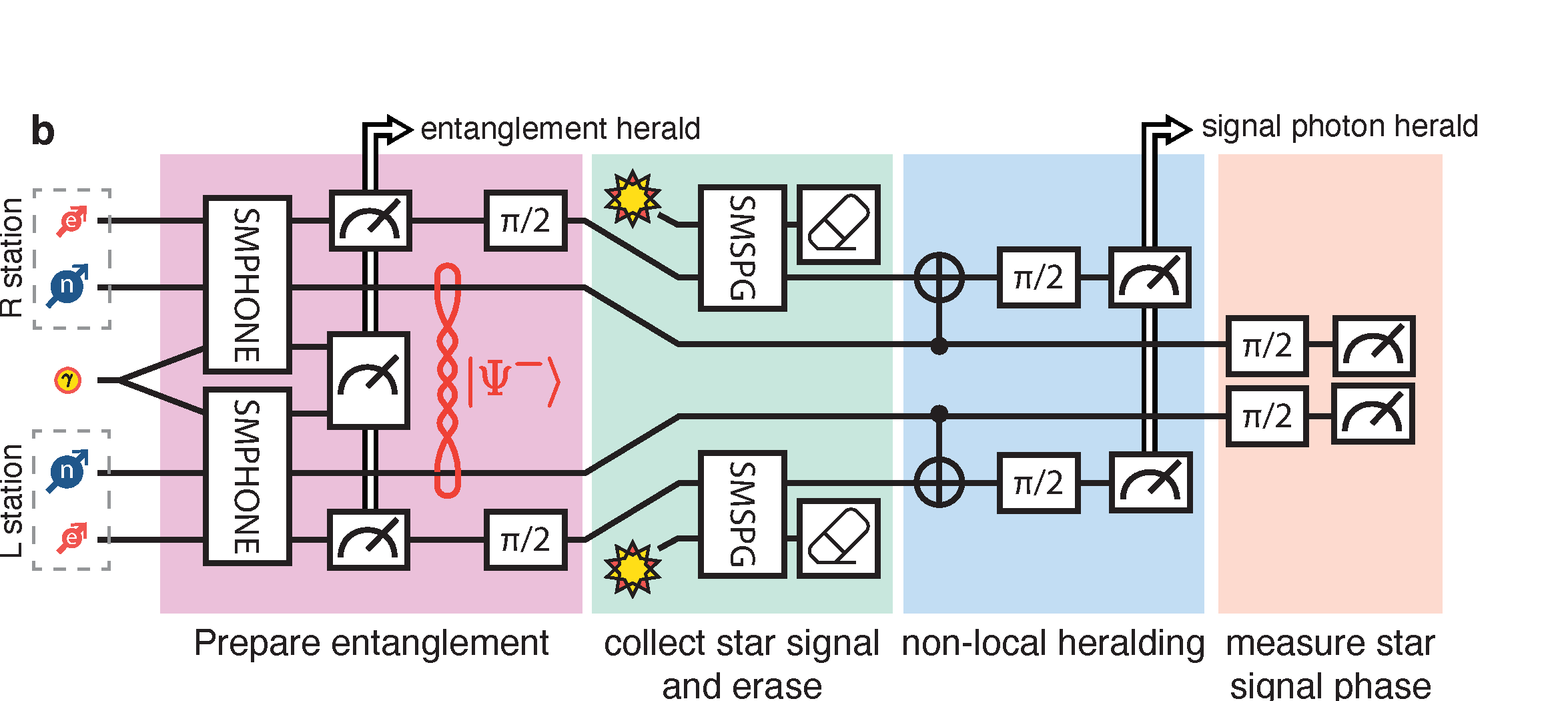}
\caption{Diamond-memory sequence: entangle the two nodes, collect the weak signal without revealing its station, report a signal-present event, and read the stored differential phase. Cropped from Fig.~4b of P.-J. Stas \etal, arXiv:2509.09464v2 (2025), under the \href{https://creativecommons.org/licenses/by/4.0/}{CC BY 4.0} license; see also the published version.\cite{Stas2026}}
\label{fig:stas}
\end{figure}

The nodes were about 6~m apart. Stas \etal\ emulated an effective fiber length $L=1.55$~km using 3.1~km of spooled fiber across the two entanglement paths. For the complete experiment with the nodes physically separated by 6~m, the average accepted-event rate was roughly one event every 80 seconds, over mean input levels from 0.25 to 2 photons per pulse. Figure~\ref{fig:stas} shows the sequence. The experiment combined several operations needed for a memory-assisted receiver; rate, false heralds, cryogenic hardware, and bandwidth remain limiting.

\subsection{Cold-atom memories and long delay compensation}

\begin{wrapfigure}[20]{R}{0.44\textwidth}
\vspace{-8pt}
\centering
\includegraphics[width=0.96\linewidth]{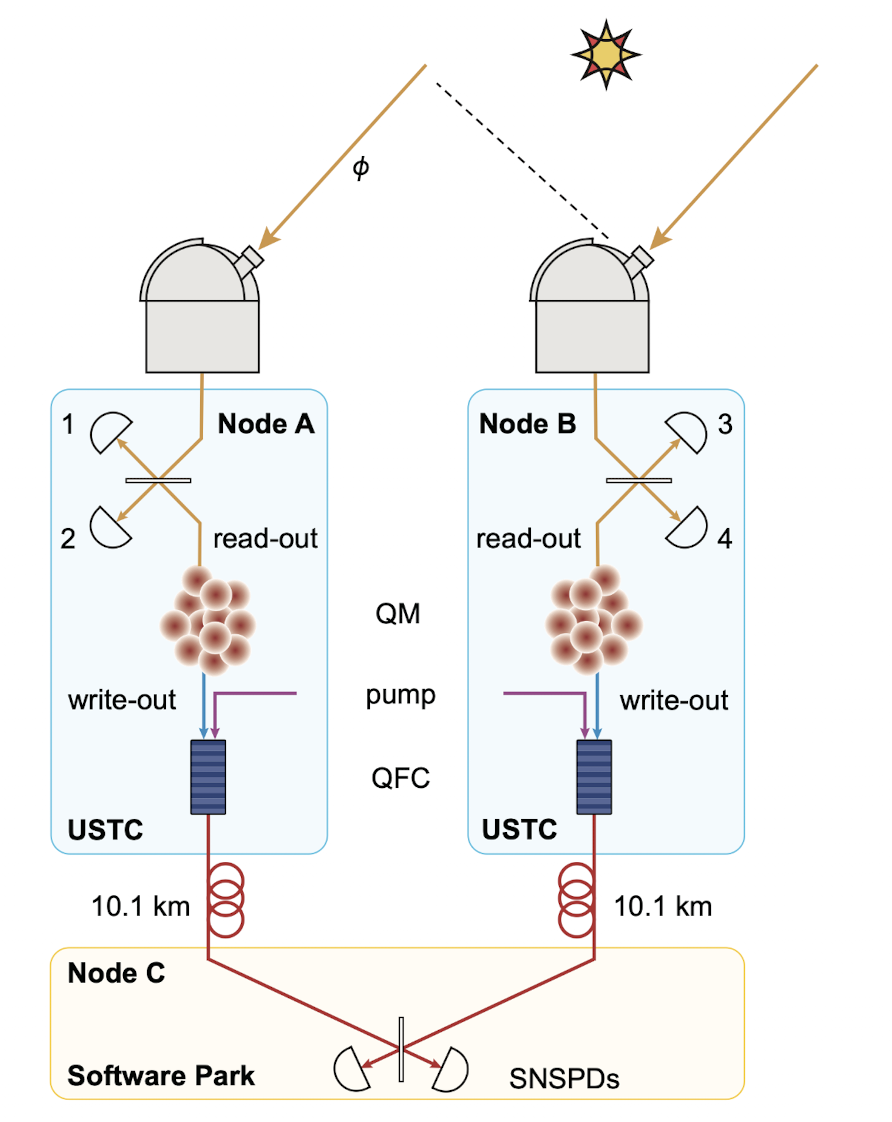}
\caption{Cold-atom memory-assisted interferometer. Nodes A and B store a shared atomic excitation; a detector click at remote node C heralds successful preparation. Reprinted with permission from Fig.~1 of \href{https://doi.org/10.1103/qpzn-h7p9}{B. Wang \etal, \textit{Phys. Rev. Lett.} \textbf{136}, 240801 (2026)}. Copyright (2026) by the American Physical Society.}
\label{fig:wang}
\vspace{-8pt}
\end{wrapfigure}

Wang \etal\ implemented a memory-assisted Brown/GJC receiver in which cold-atom memories stored and later released the shared reference resource; the simulated starlight itself was not stored. They demonstrated one programmed readout delay rather than a changing astronomical delay, and they did not implement the Khabiboulline binary time-bin protocol. They used two $^{87}$Rb ensemble memories, clouds of atoms in which an optical excitation can be held collectively by many atoms, separated by about 2~m.\cite{Wang2026} To prepare the reference, each memory was connected through a 10.1-km deployed-fiber arm to a common remote node. A detector click there heralded that one atomic excitation was stored across the two memories without determining which memory held it. When released, this excitation became the path-entangled reference photon used in the GJC measurement. At each receiver, this reference interfered with one arm of a split laboratory source that mimicked thermal starlight. The reported 20-km link is the sum of the two fiber arms, not the separation of the memories (Fig.~\ref{fig:wang}).

The local write and read times were programmed in advance. The remote herald time tag was used afterward to select successful records; the distant click did not trigger memory readout in real time after a round-trip message. The final optical measurement was similar to Brown \etal's implementation of GJC: Wang \etal\ scanned the reference phase and compared coincidence rates from the two receivers to recover the visibility amplitude and phase of the simulated source. A 1-km fiber spool and a programmed $5~\mu$s difference between the two memory readout times tested one fixed delay, equivalent to 1.5~km of free-space path.

\newpage

\subsection{Kwiat-group three-telescope tests}

The Kwiat group has explored extending the GJC architecture to three telescopes in a series of conference reports and a poster.\cite{Diaz2021,Diaz2022,Diaz2024Poster} This work considers visibility recovery and closure phase in the presence of phase turbulence. The published record is still preliminary, and a full archival experimental paper has not yet appeared.

\subsection{Two-source two-photon astrometry}

Stankus \etal\ proposed replacing GJC's engineered reference photon with light from a second astronomical source.\cite{Stankus2022} At each telescope, diffraction-limited modes from the two source directions are mixed locally. Coincidences between the stations retain the difference between the two sources' first-order interferometric phases. The measurement is therefore GJC-like rather than HBT-like, but it shares one practical advantage with HBT: no entangled resource or phase-stable optical link need be distributed between the stations. The stations instead exchange photon-detection records, including arrival times and detector channels; their clocks and geometric delays must still be aligned well enough to identify coincident pairs. However, using an astronomical source instead of an engineered reference source comes with significant trade-offs. One photon from each source must arrive within the same narrow coincidence window, so the useful rate is proportional to the product of the two small photon occupations. Further, the second source must not be fully resolved on the long baseline, since resolving it reduces the detected fringe contrast. In contrast, an engineered GJC reference source can in principle be driven at a high repetition rate, whereas the second star supplies photons randomly at a rate that the instrument cannot control. The method also retains strict GJC-like local requirements for matching spectrum, polarization, spatial mode, and differential delay; eliminating the long inter-station optical link does not eliminate delay control within each telescope. Crawford \etal\ demonstrated the predicted correlation and anticorrelation with two thermal-like laboratory sources, but not separated-station or on-sky astrometry.\cite{Crawford2023}

\WFclear
\subsection{A quantum hard drive}

Bland-Hawthorn, Sellars, and Bartholomew proposed storing the astronomical field in solid crystals doped with rare-earth ions, transporting the loaded memories, and reading them together later.\cite{BlandHawthorn2021} Such a memory could in principle store many independently addressable spatial, spectral, and temporal modes. Rare-earth materials have preserved an atomic quantum state for several hours and an optical field for one hour, although hour-scale storage and retrieval of single photons has not been demonstrated.\cite{Zhong2015,Ma2021}

The ``quantum hard drive'' remains a design study. Its 20-disk example assumes roughly $2\times10^7$ independently addressable storage cells and up to $2\times10^{10}$ optical modes. Neither that capacity nor coherent transport of a loaded memory has been demonstrated. Before they could interfere, corresponding stored modes from the two telescopes would still have to be read out with the correct relative timing and optical phase, including compensation for the geometric delay. The proposed advantage is to write, transport, and read the field instead of building a long delay line.

\FloatBarrier
Table~\ref{tab:experiments} summarizes the selected literature and its role in this review. While not comprehensive, these papers illustrate the development of the field and the experimental progress toward practical implementations.

\begin{table}[ht]
\caption{Selected milestones in quantum-enabled optical interferometry; not a comprehensive bibliography.}
\label{tab:experiments}
\centering
\footnotesize
\begin{tabularx}{\linewidth}{>{\raggedright\arraybackslash}p{0.22\linewidth} >{\raggedright\arraybackslash}p{0.13\linewidth} >{\raggedright\arraybackslash}X}
\toprule
Reference & Status & Primary contribution \\
\midrule
Gottesman, Jennewein, and Croke (2012)\cite{Gottesman2012} & Theory & Introduced a path-entangled reference photon for recovering first-order coherence without transporting starlight. \\
\addlinespace[2pt]
Khabiboulline \etal\ (2019a,b)\cite{Khabiboulline2019PRL,Khabiboulline2019PRA} & Theory & Introduced binary time-bin encoding with quantum memories, avoiding a flying reference in every bin and reducing entanglement use from linear to logarithmic; extended the method to broadband and multiple-station operation. \\
\addlinespace[2pt]
Brown \etal\ (2023)\cite{Brown2023} & Experiment & Implemented the GJC entangled-reference receiver with pseudo-thermal light, measured complex visibility at individual baselines, and used $|V|^2$ across baselines to reconstruct the intensity autocorrelation of a binary-star analogue. \\
\addlinespace[2pt]
Diaz \etal\ (2021--2024)\cite{Diaz2021,Diaz2022,Diaz2024Poster} & Conference reports / poster & Explored three-telescope extensions of GJC, with emphasis on visibility recovery and closure phase. \\
\addlinespace[2pt]
Bland-Hawthorn \etal\ (2021)\cite{BlandHawthorn2021} & Concept & Proposed a multimode rare-earth ``quantum hard drive'' for storing, transporting, and later reading the astronomical field. \\
\addlinespace[2pt]
Stankus \etal\ (2022)\cite{Stankus2022} & Theory & Used a second star as the reference for GJC-like local interference, avoiding distributed entanglement but requiring rare two-source coincidences and local mode and delay matching. \\
\addlinespace[2pt]
Stas \etal\ (2026)\cite{Stas2026} & Experiment & Demonstrated Khabiboulline-style nonlocal phase sensing with event-ready entanglement between two memory nodes, including mode erasure, nonlocal heralding, and phase readout; no binary time-bin encoding. \\
\addlinespace[2pt]
Wang \etal\ (2026)\cite{Wang2026} & Experiment & Implemented a memory-assisted GJC receiver using cold-atom memories and long fiber links, including a fixed programmed readout delay. \\
\addlinespace[2pt]
Gorshkov \etal\ (2026)\cite{Gorshkov2026PhaseClosure} & Theory / conference & Explored quantum-optimized phase closure for robust imaging with multiple telescopes. \\
\bottomrule
\end{tabularx}
\end{table}

\FloatBarrier
\section{Technological requirements}
\label{sec:technology}

A realistic sky experiment depends on the product of source coupling, network rate, memory efficiency, gate fidelity, readout, detector efficiency, and phase stability, across enough modes to collect useful starlight. Fortunately, the enthusiasm for all things quantum is leading to some true breakthroughs in the critical enabling technologies (see Table~\ref{tab:stack}).

\begin{table}[!b]
\caption{Functions that a quantum-enabled optical array would have to supply.}
\label{tab:stack}
\centering
\small
\begin{tabularx}{\linewidth}{>{\raggedright\arraybackslash}p{0.27\linewidth} X}
\toprule
Technology & Technology function \\
\midrule
Superconducting nanowire single-photon detectors and other fast photon counters & Devices that register individual photons with high efficiency, low dark-count rate, and precise timing for correlations and heralding \\
\addlinespace[2pt]
Quantum memories & Material systems that preserve a loaded quantum state until it is needed, providing a calibrated delay and a stationary resource for later measurements \\
\addlinespace[2pt]
Light--memory interfaces and quantum gates & Hardware that transfers astronomical optical modes into and out of memory, performs controlled operations on the stored states, preserves their relative phase, and works across a wide optical bandwidth \\
\addlinespace[2pt]
Entangled-photon sources and quantum repeaters & Sources that generate correlated or entangled pairs, plus intermediate nodes that extend those correlations across lossy links \\
\addlinespace[2pt]
Fiber links and quantum frequency conversion & Links between telescope nodes; frequency conversion changes the photon wavelength so low-loss telecom transport and memory or science wavelengths can work together \\
\addlinespace[2pt]
Optical clocks, frequency combs, and stabilized timing & Precise timing and optical-frequency references for synchronizing temporal modes, phases, geometric delays, and classical records \\
\addlinespace[2pt]
Photonic integrated circuits & Compact optical chips containing stable beam splitters, phase shifters, filters, and coincidence-measurement circuits that can be replicated over many channels \\
\bottomrule
\end{tabularx}
\end{table}

\subsection{Bandwidth and mode count}

Astronomers gain sensitivity by observing over broad bandwidths. Near-infrared interferometers, for example, often use the full H band from 1.5 to 1.8~$\mu$m, a window of roughly 30~THz. If each memory interface accepts 1~GHz, covering that band would require about $3\times10^4$ spectral channels before counting successive temporal modes, spatial modes, or polarization states. Each channel would need compatible conversion, timing, phase calibration, processing, and detection. Processing many channels in parallel, or multiplexing, will be essential to collect enough stellar continuum to take advantage of the long baselines that motivate these methods.

Khabiboulline \etal\ showed that an incoming photon's arrival time and frequency band can be stored efficiently in a compact binary code.\cite{Khabiboulline2019PRA} This reduces the required memory and entanglement resources, but it does not remove the need to split and interface the light across many spectral channels. The optical hardware still scales with bandwidth.

Spectral bandwidth is only one part of the mode-count problem. Atmospheric residuals continually move starlight among spatial modes, while a single-mode quantum interface rejects everything outside its acceptance mode. Polarization must also be preserved or measured. Adaptive optics, photonic lanterns, or replicated spatial channels may therefore be needed before the quantum protocol even begins.

\subsection{End-to-end efficiency and fidelity}

To keep the comparison compact, let the effective usable-resource factor be represented schematically as
\begin{equation}
    \eta_q = \eta_{\rm ent}\,\eta_{\rm couple}\,\eta_{\rm memory}\,
    \eta_{\rm gate}\,\eta_{\rm read}\,\eta_{\rm det}\,\mathcal{V}_q^2.
    \label{eq:etaq}
\end{equation}
The factors represent entanglement delivery, coupling, storage, gates, readout, and detection, while $\mathcal{V}_q$ represents phase stability and the quality of the shared quantum state. This is a bookkeeping expression, not a literal loss budget. Some imperfections remove events, while others reduce the measured contrast or introduce errors. Here fidelity means how closely the delivered quantum state matches the intended state. Six successive 80\% efficiencies leave only 26\% before the visibility penalty. If successful entanglement generation is announced before the astronomical mode is processed (``event ready''), failed attempts can be rejected; however, a low generation rate reduces the number of usable observing modes. Parallel network channels, error correction, and long-lived memories will help, but the end-to-end efficiency must be high enough to make the quantum layer worthwhile compared with direct detection.

\subsection{Time-dependent delay and atmosphere}

Earth rotation changes the geometric delay continuously, while atmospheric turbulence adds faster path-length fluctuations above each telescope. Any architecture must bring the interfering modes into temporal overlap and control their relative phase. A quantum link does not remove the need for fringe acquisition, dispersion control, or correction of atmospheric piston.

One approach is an optical delay: keep the state as a traveling photon and change the physical path it follows. This is the familiar solution in direct detection and is also required when flying reference photons or two astronomical photons must meet at local beam splitters. Compact optical buffers may replace part of a long free-space delay line. A Kwiat-group free-space memory combined three switchable optical loops to provide delays up to 12.5~$\mu$s, with 5\% end-to-end efficiency at the longest delay.\cite{Arnold2023} Guo \etal\ retrieved about 95\% of photons after a selectable 0.69-$\mu$s delay in a nested multipass cell while preserving polarization entanglement at 99.6\% fidelity.\cite{Guo2026} These devices demonstrate useful switches, coatings, and photon storage paths, but not the continuously variable, phase-referenced delay needed to follow a star.

Quantum memories offer a more elegant approach for a quantum network: map the optical state into a stationary material excitation, prepare and herald network resources in advance, and release or process the state when the required delay is reached. The geometric delay can then be handled partly through programmed storage and readout rather than only through meters or kilometers of propagation. Important elements have now been demonstrated: Stas \etal\ used event-ready entanglement between remote memories for nonlocal phase sensing, while Wang \etal\ stored a distributed reference and programmed a fixed $5~\mu$s difference between two memory readout times, equivalent to 1.5~km of free-space path.\cite{Stas2026,Wang2026} Neither experiment continuously tracked an astronomical delay. An astronomical implementation would still need efficient broadband loading, many independent modes, phase-preserving storage, and readout times that follow the changing delay. Quantum memory therefore provides an attractive way to manage delay within a quantum network, even though it moves the precision control into the write, storage, and readout system rather than eliminating it.

\subsection{Scaling beyond a baseline}

An $N$-telescope array has $N(N-1)/2$ baselines and $(N-1)(N-2)/2$ independent closure phases. Entangling telescope pairs, sharing one state among many telescopes, or using a central device that dynamically routes entanglement connections (a quantum switch) leads to different routing, clock, scheduling, and calibration requirements. Array tests must address simultaneous baselines, unequal node losses, and station-dependent phases.

\section{Where can quantum interferometry add value?}
\label{sec:competition}

In the astronomical interferometry context, where the light is weak, broadband, and emitted incoherently, directly combining the light is, in the ideal lossless case, the most sensitive way to measure its coherence.\cite{Tsang2011} Any practical advantage of quantum methods must therefore come from how that measurement is implemented, including its loss, cost, and complexity. I would compare each quantum measurement with the best classical alternative using the same apertures and incident modes.

\paragraph{Improved ground-based direct detection.}
Low-loss beam transport (free-space or fiber), better coatings, adaptive optics, integrated optics combiners, or a different array geometry may save more photons than a quantum layer. These approaches preserve the first-order measurement without the network overhead or the problem of multiplexing thousands of parallel modes. In terms of sensitivity and imaging, I expect direct detection to remain best for ground arrays up to a few km for the foreseeable future.

\paragraph{Intensity interferometry and heterodyne detection.}
Some astronomers are pushing HBT and heterodyne techniques to longer baselines, new wavelengths, and more telescopes, ironically driven by technologies developed for quantum experiments.
For HBT, coupling a high-resolution photonic spectrograph to arrays of single-photon avalanche diodes (SPADs) or fast superconducting nanowire detectors enables many narrow spectral channels to be correlated in parallel. Photonic spectrographs can also multiplex heterodyne channels, while frequency combs provide phase-referenced local oscillators across them. Cherenkov telescope arrays have been studied as kilometer-scale HBT interferometers, including the Cherenkov Telescope Array.\cite{Dravins2013CTA} The IC4Stars project aims to measure Sirius~B with two 8-m-class telescopes separated by about 1~km, using many wavelength channels to increase sensitivity. Its recent preprint reports a single-telescope on-sky test with the 1-m Epsilon telescope at Calern; the long-baseline measurement remains future work.\cite{IzraelevitchPatitucci2026} If the goal is to observe very long baselines without transporting starlight, these classical methods may be the most practical way to do it in the near future, at least for selected bright targets.

\paragraph{Space interferometry.}
Space provides low-loss propagation, long baselines, no atmospheric piston, and lower infrared background for a cryogenic interferometer array. The costs are driven by collecting area, formation control, pointing, timing, and the intrinsic risk of the mission itself. Reconfiguring the baselines also requires propulsion and consumes propellant, potentially limiting both the available array configurations and the mission lifetime. Quantum hardware removes none of these constraints, although a hybrid quantum-space array may be useful.

\paragraph{Microarcsecond astronomy.}
Multi-kilometer optical/infrared baselines reach angular scales unavailable by other means, and differential astrometry can be finer than $\lambda/B$ with stable calibration. Applications could include exoplanet systems and compact emission around supermassive black holes. Unfortunately, bright stars are also resolved on long baselines, weakening their coherence and usefulness as astrometric references, and atmospheric turbulence\cite{shaocolavita} further degrades the fundamental limit unless objects are quite close together on the sky. A quantum link does not remove these constraints, although it may allow longer baselines than practical direct detection.

\paragraph{Measurements beyond astronomical imaging.}
Bright engineered beacons may provide a more accessible near-term application than weak broadband starlight. At $1~\mu$m, a 10-km baseline corresponds to an angular scale of about $20~\mu\mathrm{as}$. A beacon at a known location could serve as a bright test source for array phasing and baseline calibration, with possible applications to metrology or geodesy.
New ultra-precise methods could potentially be applied to gravitational wave detection or new tests of general relativity.

\section{Next experiments}
\label{sec:roadmap}

The next leap for quantum interferometry will be to move out of the laboratory and into an observatory dome to measure real thermal starlight. A first on-sky visibility measurement would expose the full observatory problem, including coupling into the required modes, atmospheric piston, delay tracking, and calibration. Work on longer links, higher rates, better memories, multimode operation, and new platforms can proceed in parallel. Some specific tests follow:

\begin{enumerate}[leftmargin=*,itemsep=3pt]
    \item \textbf{Can we recover a visibility on the sky?} Apply an entangled-reference or memory-assisted protocol first to spatially filtered sunlight as a receiver test and then to a bright star. Does the calibrated stellar visibility agree with conventional direct detection?
    \item \textbf{What are the practical challenges of separate telescope operation?} Place nodes at physically separated telescopes and track geometric delay and piston while preserving nonlocal heralding and path erasure.
    \item \textbf{Can we use enough modes?} Operate many spectral channels simultaneously and report sensitivity per incident astronomical mode, including coupling, conversion, memory, network, gate, and detector losses.
    \item \textbf{Can three stations recover closure phase?} Measure closure phase on a thermal source for simple image reconstruction.
    \item \textbf{Does a complete quantum experiment win in a science measurement?} Choose a narrow case, perhaps stellar-diameter visibility, binary astrometry, or geodesy, and compare observing time and systematic error against HBT, heterodyne, and direct detection using the same apertures. It is conceivable that quantum methods could become competitive with heterodyne or HBT in the near future.
    \item \textbf{Build the next interferometer to be ``quantum-ready.''} As astronomers build new direct-detection facilities, the infrastructure should include fiber routes, laboratory and cryogenic space near telescope nodes, access to clocks and frequency combs, calibration paths, and time-tagged photon streams. We lack a clear open platform for quantum physicists to test new ideas, and astronomers lack a clear path to test quantum methods on real starlight. A shared facility would benefit both communities.
    \item \textbf{Continue brainstorming about new applications.} Quantum interferometry represents a fundamentally different way to measure the coherence of light over potentially huge distances. It is too early to tell how it will be used, and the highest-impact applications may not be astronomical. The community should continue to explore new ideas, new quantum resources, and new measurement goals.
\end{enumerate}

\section{Conclusions}
\label{sec:conclusions}

After more than a decade of theoretical proposals, quantum-assisted optical interferometry has finally become a laboratory reality. Recent experiments have recovered weak-light spatial coherence with entangled references, heralded and read out a phase shared between separated memory nodes, and recovered complex visibility with a memory-assisted reference prepared over deployed-fiber links, including a test with a fixed differential delay. None has yet reported first observations with astronomical light\footnote{Rumor has it first fringes on the Sun have been obtained as I am writing this\ldots}. Bandwidth, event rate, wavelength coverage, scaling to more than two telescopes, and delay tracking remain major practical challenges.

Ideal direct detection remains the benchmark. Entanglement supplies no source photons, but a quantum resource can be regenerated, heralded, stored, or routed. This may extend first-order interferometry beyond the practical reach of classical beam transport. Experiments must determine where the efficiency, rate, stability, and cost are favorable.

The next decade should see a flurry of experiments with real starlight, exposing constraints absent from laboratory sources as quantum technologies continue to advance. Baselines beyond 10~km may also motivate ideas that previously seemed impractical, including tests with engineered beacons from Earth orbit outward. Astronomers and observatories should remain open to hosting well-designed experiments on large telescopes and interferometers. This moment recalls the experiments my PhD advisor, Charles Townes, pursued in the 1970s and 1980s as advances in lasers and nonlinear optics crossed into astronomy. That history is a useful reminder to take new technical opportunities seriously, even before their best applications are obvious.

\begin{acknowledgments}
The author thanks Michael G. Raymer, Matthew R. Brown, Brian J. Smith, Stephen T. Ridgway, Paul G. Kwiat, Pieter Stas, and Emil T. Khabiboulline for helpful discussions and contributions that informed this review. J.D.M. acknowledges support from the U.S. National Science Foundation under Award No.~1936321, \textit{QII-TAQS: Quantum-Enhanced Telescopy} (PI: Paul G. Kwiat, University of Illinois Urbana-Champaign; J.D.M. co-PI), and Award No.~2326803, \textit{QuSeC-TAQS: Distributed Entangled Quantum-Enhanced Interferometric Imaging for Telescopy and Metrology} (PI: Brian J. Smith, University of Oregon; J.D.M. co-PI).

OpenAI's ChatGPT was used to assist with manuscript organization, language editing, and preparation of figure and source inventories. The author reviewed and takes responsibility for the scientific content, interpretations, citations, and final text.
\end{acknowledgments}

{\footnotesize
\setlength{\baselineskip}{8.5pt}
\bibliography{SPIE_2026_Quantum_Interferometry_Review}
\bibliographystyle{spiebib}
}

\end{document}